# Low Temperature Relaxation of Donor Bound Electron Spins in $^{28}$Si:P

E. Sauter,[1] N. V. Abrosimov,[2] J. Hübner,[1] and M. Oestreich[1,*]

[1]*Institut für Festkörperphysik, Leibniz Universität Hannover, 30167 Hannover, Germany*
[2]*Leibniz-Institut für Kristallzüchtung, 12489 Berlin, Germany*

We measure the spin-lattice relaxation of donor bound electrons in ultrapure, isotopically enriched, phosphorus-doped $^{28}$Si:P. The optical pump-probe experiments reveal at low temperatures extremely long spin relaxation times which exceed 20 h. The $^{28}$Si:P spin relaxation rate increases linearly with temperature in the regime below 1 K and shows a distinct transition to a $T^9$ dependence which dominates the spin relaxation between 2 and 4 K at low magnetic fields. The $T^7$ dependence reported for natural silicon is absent. At high magnetic fields, the spin relaxation is dominated by the magnetic field dependent single phonon spin relaxation process. This process is well documented for natural silicon at finite temperatures but the $^{28}$Si:P measurements validate additionally that the bosonic phonon distribution leads at very low temperatures to a deviation from the linear temperature dependence of $\Gamma$ as predicted by theory.

Silicon is one of the most promising materials for spin-based quantum information processing [1–12] in which the increasing availability of isotopically purified $^{28}$Si [13] accelerates the development of prospective spin-based quantum information technologies. Recent experiments on phosphorus donor bound electrons in $^{28}$Si:P have shown very efficient optical addressability [14], extremely high fidelity qubits [3], and electronically tunable qubit gates [15]. The foundation for this success is the excellent preservation of the phase of the two level donor electron spin superposition which is described by the coherence time $T_2$. In natural silicon, $T_2$ is at low temperatures mainly limited by the interaction of the electron spin with the $^{29}$Si nuclear spins [16]. In isotopically enriched $^{28}$Si, the reduction of the number of nuclear spins reduces the associated magnetic noise and the coherence time becomes very long. So far, spin coherence times exceeding seconds have been experimentally demonstrated [3] but the fundamental limit is finally defined by the electron spin relaxation time $T_1$ with $T_2 \leq 2T_1$. The spin-lattice relaxation time $T_1$ or rather the spin relaxation rate $\Gamma = 1/T_1$ of donor bound electrons ($D^0$) has been extensively studied in phosphorus-doped natural silicon decades ago [17–23]. The experiments have been carried out for temperatures ranging from 0.4 K up to room temperature in dependence on magnetic field and yield for temperatures well below the Debye temperature of silicon ($\approx 645$ K) the empirical equation [19,24]

$$\Gamma = k_H H^4 T + k_7 H^2 T^7 + k_9 T^9 + E(H)e^{-\Delta/k_B T}, \quad (1)$$

where $k_B$ is the Boltzmann constant, the other $k_i$ are empirical constants, $H$ is the magnetic field, $T$ is the lattice temperature, and $\Delta$ is the energy difference between the ground and the first relevant excited state of the donor bound electrons. The term linear in temperature is attributed to single phonon absorption and emission [25] which require a phonon energy of $g\mu_B H$, where $g \approx 2$ is the $D^0$ Landé $g$ factor. The next two terms are attributed to two-phonon Raman scattering processes where the $T^7$ term is interpreted as a magnetic type of interaction and the $T^9$ as spin-orbit interaction [25]. The exponential term results from phonon induced transitions from the donor ground state to an excited state which is known as the Orbach process [26] and becomes relevant at $T \gtrsim 7$ K in silicon [27].

Several experiments [21–23] show that the spin lattice relaxation at low temperatures is also influenced by fast relaxation centers which depend on donor concentration and the degree of compensation of donors by acceptors. The latter effect especially is not fully understood, and makes the experimental results strongly dependent on sample quality. In fact, the hitherto existing temperature dependent measurements of $\Gamma$ neither show a clear $T^7$ nor a clear $T^9$ dependence between 2.5 and 6 K, but in general something in between. Theoretical investigations concerning these nondistinctive results point out that the calculations of the respective Raman processes are rather difficult [21]. The situation is further complicated by the nuclear spin of the $^{29}$Si isotope in natural silicon which sets the stage for nuclear spin diffusion. As a consequence, reliable predictions of the low temperature limit of $\Gamma$ in isotopically enriched $^{28}$Si are lacking.

In this Letter, we study the temperature and magnetic field dependence of $\Gamma$ in ultrapure $^{28}$Si refined from the Avogadro project [28]. The measurements cover the temperature range from 0.5 to 4.0 K revealing all the $T_1$ spin physics relevant for $D^0$ spin qubit operations. The $4 \times 2 \times 0.8$ mm$^3$ bulk $^{28}$Si sample is isotopically enriched to 99.995%, $n$ doped with a nominal phosphorus doping concentration of $n_d = 1.2 \times 10^{15}$ cm$^{-3}$ and has an acceptor background concentration $n_a < 10^{12}$ cm$^{-3}$. The sample is placed fully strain-free in a specifically designed helium gas reservoir coupled to the cold finger of a cryogen-free dilution refrigerator. We measure $\Gamma$ by an optical spin-pumping pump-probe scanning method based on Ref. [29], which overcomes two of the difficulties faced by traditional electron spin resonance methods. Unlike cavity microwave resonance methods, there is no restriction on the magnetic field, which is constant over the whole measurement time, and optical pumping of the spin polarization to $\geq 80\%$ is easily possible at any given temperature and magnetic field. The optical procedure relies on the spectrally sharp donor bound exciton ($D^0X$) transitions in high-purity $^{28}$Si:P, which are pumped and probed by linearly polarized laser light propagating along the [001] crystal axis [14,30]. We apply a transverse magnetic field between 60 mT and 1.1 T along the [110] direction, which splits the coupled spin-1/2 ground states of $D^0$ and phosphorus nucleus according to the Breit-Rabi equation into four levels separated in energy. In addition, the optically excited $D^0X$ state is split into four levels due to the Zeemann splitting of the trion's spin-3/2 hole. The resulting spectrum consists of twelve dipole allowed transitions, i.e., six doublet transitions each split by half of the phosphorus donor hyperfine coupling constant $A = 117.53$ MHz, which is smaller than the linewidth of the optical transitions [30]. The inset of Fig. 1(b) illustrates schematically the corresponding energy diagram where the black arrows depict the optically allowed doublet transitions for linearly polarized light, which is used for the $D^0$ spin initialization ($\pi_-$) and the measurement of the temporal decay of the $D^0$ polarization ($\pi_-$ and $\pi_+$).

The $D^0$ are spin polarized to 80% by pumping the spectrally overlapping doublet $\pi_-$ transition $m_e^{-1/2} \to m_h^{-1/2}$ for 1 s [30] using a frequency stabilized, external cavity diode laser with a power of 1 mW. The pump laser is switched off after 1 s and the subsequent temporal decay of the induced $D^0$ spin polarization is interruptedly measured by a weak probe laser with a power of 1 $\mu$W. In practice, pump and probe light come from the same laser and have identical optical paths and identical Gaussian spot diameters in the center of the sample of $\approx 90$ $\mu$m ensuring perfect overlap between pump and probe volume. The probe light is frequency modulated using an electro-optical modulator with a modulation frequency of 30 MHz where the center frequency of the laser is swept linearly twice within 100 ms back and forth over the energy range of the relevant $D^0X$ transitions [29]. Figure 1(a) shows a typical example of the resulting differential $D^0$ absorption spectrum. The observed line shapes are in very good approximation the first derivative of a Lorentzian function. The depicted differential absorption areas $A_+$ and $A_-$ correspond to the $\pi_+$ and $\pi_-$ transitions, respectively, and yield the $D^0$ spin polarization according to $P = (A_+ - A_-)/(A_+ + A_-)$.

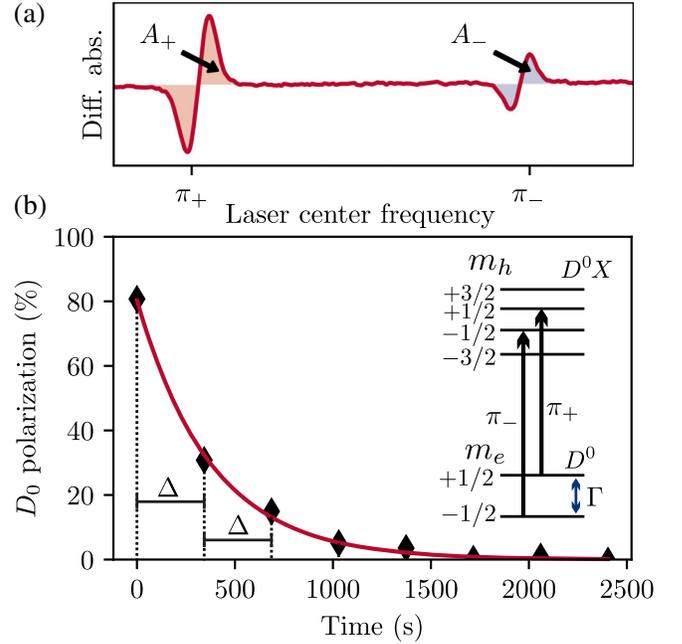

FIG. 1. (a) Typical differential absorption spectrum yielding the $D_0$ spin polarization. (b) Typical, temporal decay of the $D_0$ spin polarization versus time. The black diamonds depict the measured spin polarizations and the red line is an exponential fit according to Eq. (2). The inset represents a simplified energy diagram neglecting the phosphorus nuclear spin. The depicted measurements are carried out at a temperature and magnetic field of 3 K and 60 mT, respectively. Initialization is accomplished by optical pumping the $\pi_-$ transition which is at this magnetic field at 278.033 51(1) THz.

In order to measure the temporal dynamics of the $D^0$ spin relaxation with the lowest possible perturbation, the probe laser is blocked repeatedly for varying time intervals $\Delta$ during which the $D^0$ spins relax undisturbed by any light [31]. Figure 1(b) shows exemplarily the measured exponential decay of the $D^0$ polarization with time for $\Delta = 350$ s. Measurements with different $\Delta$ show that despite the very low light intensity and short measurement time, each of the optical absorption measurements changes the polarization of $D^0$ slightly by a factor $\lambda_L \lesssim 1$ because the sweeping laser still moves some population between the polarized states. The polarization $P$ after $n$ scans is therefore

$$P_n = \lambda_L^n e^{-\Gamma n \Delta'},$$

with $\Delta' = \Delta + 100$ ms and the intrinsic spin relaxation rate $\Gamma$. The measured transient polarization becomes

$$P(t) = P_0 e^{-\Gamma t} \lambda_L^{t/\Delta'} = P_0 e^{-(\Gamma - \ln(\lambda_L)/\Delta') \cdot t} = P_0 \exp^{-\Gamma_{\text{eff}} \cdot t}, \quad (2)$$

with the effective relaxation rate

$$\Gamma_{\text{eff}} = \left(\Gamma - \frac{\ln(\lambda_L)}{\Delta'}\right). \quad (3)$$

In order to measure the intrinsic $\Gamma$ with highest accuracy, we repeat the measurement with increasing $\Delta$ and extract $\Gamma$ according to Eq. (3). Performing the measurement this way eliminates any perturbations by the laser light and enables accurate measurements of the intrinsic $\Gamma$ at low and very low temperatures with high efficiency.

Figure 2 depicts the resulting magnetic field dependence of $\Gamma$ between 120 mT and 1.1 T in the low temperature regime from 2.5 to 3.6 K. The experimental data (black dots) show a strong increase of $\Gamma$ with increasing magnetic field and increasing temperature, which is perfectly described in this temperature regime by

$$\Gamma = k_H H^4 T + k_9 T^9. \quad (4)$$

In order to compare the coefficients $k_H$ and $k_9$ to the findings in natural silicon, we extract the relevant published data from Refs. [17–19,32]. We evaluate for this comparison only data from low doped silicon and include temperatures up to 4 K where the Orbach mechanism can still be neglected. Table I summarizes the corresponding coefficients. The electron spin resonance measurements on natural silicon are concerning $k_9$ in excellent agreement with our high accuracy optical measurements on $^{28}$Si. Concerning the magnetic field dependent $k_H$ term, our measurements have a much higher accuracy while the published measurements on natural silicon vary to some extent but within the uncertainty of the natural silicon data also the $k_H$ measurements are in accordance. Such an agreement is not surprising since neither $k_9$ nor $k_H$ depends on sample quality or the concentration of $^{29}$Si.

Next, we study the temperature and magnetic field dependence of $\Gamma$ at very low temperatures and discuss first the temperature dependence at 60 mT which is shown in Fig. 3 as a solid blue line. The direct phonon spin relaxation mechanism with $\Gamma = k_H H^4 T$ is at 60 mT and temperatures $\leq 4$ K lower than $10^{-7}$ Hz and therefore in any respect negligible. The measurements show instead at $T = 500$ mK a $D^0$ spin relaxation rate of $1.8 \times 10^{-5}$ Hz ($T_1 = 15.8$ h) which increases at low temperatures linearly with temperature as $k_1 T$ with $k_1 = 11$ $\mu$Hz K$^{-1}$. A linear extrapolation to 0 K yields a spin relaxation rate $k_0$ of 13 $\mu$Hz which translates into an extremely long $T_1$ spin relaxation time of 21 h. Even at 1.8 K, $T_1$ is as long as $\approx 5$ h, which is an order of magnitude longer than the fundamental $T_1$ limit assumed in a recent publication on $D^0$ spin coherence in $^{28}$Si:P [3]. Both terms $k_0$ and $k_1$ are probably caused by a neutral donor pair relaxation mechanism, which has been discussed, for example, in Ref. [21], and remaining unintentional impurities [33]. But the most intriguing observation is that the linear increase of $\Gamma$ with temperature turns directly into a strict $T^9$ dependence. The $T^7$ dependence, which has been discussed in previous publications on $\Gamma$ in natural silicon, can be explicitly excluded from our $^{28}$Si:P measurements and is probably related to sample purity.

Finally, we are discussing the temperature dependence of $\Gamma$ at *very* low temperatures and high magnetic fields where the magnetic-field dependent single phonon spin relaxation process dominates. Hasegawa [34] points out in his calculations that the usual $k_H H^4 T$ dependence is only valid if $g\mu_B B/k_B T \ll 1$. Replacing the approximate expression with the exact phonon distribution function [34] leads to the full expression for the spin relaxation at low and very low temperatures

$$\Gamma = k_0 + k_1 T + k_9 T^9 + k_H \frac{g\mu_B}{2k_B} H^5 \left(\frac{2}{\exp(\frac{g\mu_B H}{k_B T}) - 1} + 1\right). \quad (5)$$

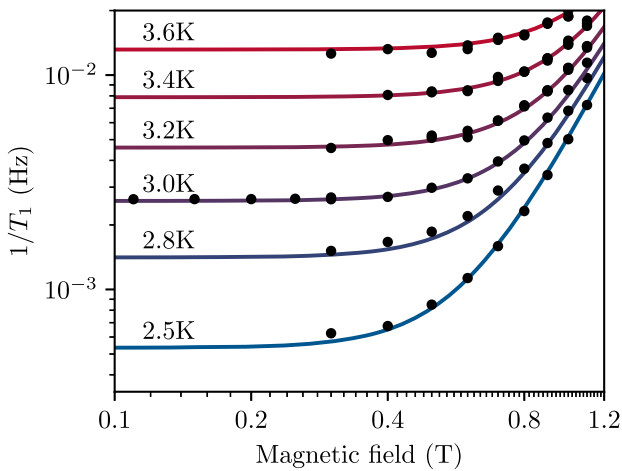

FIG. 2. Measured magnetic field dependence of $\Gamma = 1/T_1$ (black dots) in the low temperature regime. Solid lines are calculated according to Eq. (5), which yields in this regime quasi-identical results as the approximation $\Gamma = k_H H^4 T + k_9 T^9$.

TABLE I. Coefficients in Eq. (5) used to describe the measured magnetic field and temperature dependence of $\Gamma$ (top row) in comparison to published data measured in natural silicon (two lower rows).

| $k_0$ ($\mu$Hz) | $k_1$ ($\mu$Hz K$^{-1}$) | $k_H$ (mHz T$^{-4}$ K$^{-1}$) | $k_9$ ($\mu$Hz K$^{-9}$) |
|---|---|---|---|
| 13(2) | 11(3) | 1.8(1) | 0.13(2) |
|  |  | 2.6[a] | 0.1[c] |
|  |  | 0.8[b] | 0.13[d] |

[a]Ref. [17]$n_d = 1 \times 10^{15}$ cm$^{-3}$ with $H^4$ fit from extracted data.
[b]Ref. [32]$n_d = 1 \times 10^{15}$ cm$^{-3}$ with $H^4$ fit from extracted data.
[c]Ref. [19]$n_d = 9 \times 10^{15}$ cm$^{-3}$.
[d]Ref. [18]$n_d = 7 \times 10^{15}$ cm$^{-3}$ with $T^9$ fit from extracted data.

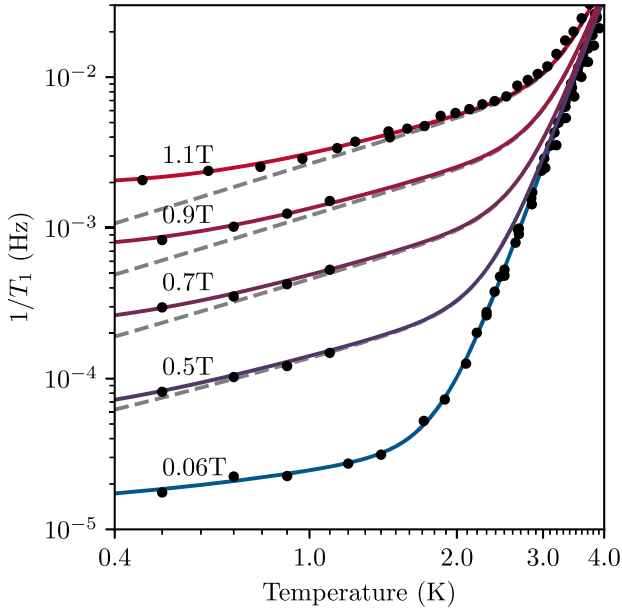

FIG. 3. Measured temperature dependence of $\Gamma$ (dots) for different magnetic fields. Solid lines are calculated according to Eq. (5) and the dashed lines according to Eq. (4) using the coefficients in Table I. The spin relaxation rate at very low temperatures is in all cases only weakly dependent on temperature but the underlying spin relaxation mechanisms at $B = 0.06$ and $B \geq 0.5$ T differ significantly.

All solid lines in Figs. 2 and 3 are calculated by Eq. (5) with the coefficients given in Table I and yield perfect agreement with the experimental data. For comparison, the dashed lines in Fig. 3 are calculated for $B \geq 0.5$ T by Eq. (4) which is in this parameter regime equivalent to Eq. (5) with the last term approximated by $k_H H^4 T$. The significant discrepancy to the temperature dependent measurements proves that the usual linear relation between $\Gamma$ and $T$ of this spin relaxation mechanism is only a good approximation at moderate temperatures and magnetic fields, but becomes void at finite magnetic fields and very low temperatures—as predicted by theory.

The magnetic field dependence of $\Gamma$ is at low temperatures according to Eq. (5) proportional to $H^5$. Such a dependence is relevant for Si:P devices and has been observed by Morello et al. in a Si:P $D^0$, single-shot readout, double-quantum dot system [35]. They observed at a lattice temperature of 40 mK and an electron temperature of 200 mK in two different devices $\Gamma = K_{0A} + K_{5A} B^5$ and $\Gamma = K_{5B} B^5$ with $K_{0A} = 1.84$ Hz, $K_{5A} = 0.0076$ Hz T$^{-5}$, and $K_{5B} = 0.015$ Hz T$^{-5}$. This translates to $k_H = 11$ mHz K$^{-1}$ T$^{-4}$ and $k_H = 22$ mHz K$^{-1}$ T$^{-4}$, respectively. They also attribute the observed spin-lattice relaxation to the valley repopulation calculated by Hasegawa [34] but extract a single phonon coefficient $k_H$ that is about one order of magnitude larger than in bulk $^{28}$Si:P (see Table I). Such a significant difference suggests that the spin relaxation in these devices is strongly influenced by the Purcell and proximity effects where the bulk spin relaxation measured in this Letter probably sets the general lower limit for $\Gamma$ in this material system.

In conclusion, we establish an optical technique of spin pumping and low disturbance, scanning absorption measurements in $^{28}$Si as a very efficient and accurate low temperature alternative to spin relaxation measurements by electron spin resonance. The optical technique yields in isotopically enriched, high quality $^{28}$Si not only the coefficient for the magnetic field dependent single phonon process but also for the $T^9$ two-phonon Raman process with superior accuracy serving as future reference for $T_1$ spin relaxation of phosphorus bound electrons in silicon. The high accuracy of the temperature dependent measurements at low magnetic field also exclude the possibility of a $T^7$ Raman process which could neither reliably be ruled out nor confirmed from the available measurements on natural silicon. The spin relaxation times at low magnetic fields and low temperatures become very long, which is interesting in view of spin based $^{28}$Si:P quantum information devices. However, some experiments on natural silicon report similarly long spin relaxation times demonstrating that $T_1$ does not depend on the nuclear spin concentration, in contrast to $T_2$ [3]. There has been a long predicted deviation from the linear relation of $\Gamma$ and $T$ for very low temperatures and high magnetic fields ($k_B T \lesssim g \mu_B H$) [34]. This prediction has been unambiguously attested by experiment for the first time.

We thank M. Glazov and M. Durnev for helpful discussions. This work was funded by the Deutsche Forschungsgemeinschaft (DFG, German Research Foundation) under Germany's Excellence Strategy - EXC-2123 QuantumFrontiers - 390837967, research training group 1991, and OE 177/10-2.

[*] oest@nano.uni-hannover.de
[1] B. E. Kane, A silicon-based nuclear spin quantum computer, Nature (London) **393**, 133 (1998).
[2] J. J. Pla, K. Y. Tan, J. P. Dehollain, W. H. Lim, J. J. Morton, D. N. Jamieson, A. S. Dzurak, and A. Morello, A single-atom electron spin qubit in silicon, Nature (London) **489**, 541 (2012).
[3] A. M. Tyryshkin, S. Tojo, J. J. L. Morton, H. Riemann, N. V. Abrosimov, P. Becker, H.-J. Pohl, T. Schenkel, M. L. W. Thewalt, K. M. Itoh, and S. A. Lyon, Electron spin coherence exceeding seconds in high-purity silicon, Nat. Mater. **11**, 143 (2012).
[4] F. A. Zwanenburg, A. S. Dzurak, A. Morello, M. Y. Simmons, L. C. L. Hollenberg, G. Klimeck, S. Rogge, S. N. Coppersmith, and M. A. Eriksson, Silicon quantum electronics, Rev. Mod. Phys. **85**, 961 (2013).
[5] K. Saeedi, S. Simmons, J. Z. Salvail, P. Dluhy, H. Riemann, N. V. Abrosimov, P. Becker, H.-J. Pohl, J. J. Morton, and M. L. Thewalt, Room-temperature quantum bit storage

exceeding 39 minutes using ionized donors in silicon-28, Science **342**, 830 (2013).

[6] M. A. Broome, T. F. Watson, D. Keith, S. K. Gorman, M. G. House, J. G. Keizer, S. J. Hile, W. Baker, and M. Y. Simmons, High-Fidelity Single-Shot Singlet-Triplet Readout of Precision-Placed Donors in Silicon, Phys. Rev. Lett. **119**, 046802 (2017).

[7] D. Keith, M. G. House, M. B. Donnelly, T. F. Watson, B. Weber, and M. Y. Simmons, Single-Shot Spin Readout in Semiconductors Near the Shot-Noise Sensitivity Limit, Phys. Rev. X **9**, 041003 (2019).

[8] Y. He, S. K. Gorman, D. Keith, L. Kranz, J. G. Keizer, and M. Y. Simmons, A two-qubit gate between phosphorus donor electrons in silicon, Nature (London) **571**, 371 (2019).

[9] E. Ferraro and E. Prati, Is all-electrical silicon quantum computing feasible in the long term?, Phys. Lett. A **384**, 126352 (2020).

[10] C. H. Yang, R. C. Leon, J. C. Hwang, A. Saraiva, T. Tanttu, W. Huang, J. Camirand Lemyre, K. W. Chan, K. Y. Tan, F. E. Hudson, K. M. Itoh, A. Morello, M. Pioro-Ladrière, A. Laucht, and A. S. Dzurak, Operation of a silicon quantum processor unit cell above one kelvin, Nature (London) **580**, 350 (2020).

[11] T. Kobayashi, J. Salfi, C. Chua, J. van der Heijden, M. G. House, D. Culcer, W. D. Hutchison, B. C. Johnson, J. C. McCallum, H. Riemann, N. V. Abrosimov, P. Becker, H. J. Pohl, M. Y. Simmons, and S. Rogge, Engineering long spin coherence times of spinorbit qubits in silicon, Nat. Mater. **20**, 38 (2021).

[12] L. Petit, H. G. Eenink, M. Russ, W. I. Lawrie, N. W. Hendrickx, S. G. Philips, J. S. Clarke, L. M. Vandersypen, and M. Veldhorst, Universal quantum logic in hot silicon qubits, Nature (London) **580**, 355 (2020).

[13] N. Abrosimov, D. ArefEv, P. Becker, H. Bettin, A. Bulanov, M. Churbanov, S. Filimonov, V. Gavva, O. Godisov, A. Gusev *et al.*, A new generation of 99.999% enriched 28-si single crystals for the determination of avogadros constant, Metrologia **54**, 599 (2017).

[14] M. Beck, N. V. Abrosimov, J. Hübner, and M. Oestreich, Impact of optically induced carriers on the spin relaxation of localized electron spins in isotopically enriched silicon, Phys. Rev. B **99**, 245201 (2019).

[15] Y. Wang, A. Tankasala, L. C. Hollenberg, G. Klimeck, M. Y. Simmons, and R. Rahman, Highly tunable exchange in donor qubits in silicon, npj Quantum Mater. **2**, 16008 (2016).

[16] W. M. Witzel, M. S. Carroll, A. Morello, L. Cywiński, and S. Das Sarma, Electron Spin Decoherence in Isotope-Enriched Silicon, Phys. Rev. Lett. **105**, 187602 (2010).

[17] A. Honig and E. Stupp, Electron spin-lattice relaxation in phosphorus-doped silicon, Phys. Rev. **117**, 69 (1960).

[18] G. Feher and E. A. Gere, Electron Spin Resonance Experiments on Donors in Silicon. II. Electron Spin Relaxation Effects, Phys. Rev. **114**, 1245 (1959).

[19] T. G. Castner, Raman spin-lattice relaxation of shallow donors in silicon, Phys. Rev. **130**, 58 (1963).

[20] T. G. Castner, Orbach spin-lattice relaxation of shallow donors in silicon, Phys. Rev. **155**, 816 (1967).

[21] G. Yang and A. Honig, Concentration- and compensation-dependent spin-lattice relaxation in n-type silicon, Phys. Rev. **168**, 271 (1968).

[22] J. Marko and A. Honig, Measurements of concentration-dependent spin-lattice relaxation times in phosphorus-doped silicon at low temperatures, Solid State Commun. **8**, 1639 (1970).

[23] P. R. Cullis and J. R. Marko, Electron paramagnetic resonance properties of n-type silicon in the intermediate impurity-concentration range, Phys. Rev. B **11**, 4184 (1975).

[24] Equation (2) in Ref. [19] contains an additional term proportional to $T^{13}$. This term is not included in Eq. (1) since calculations showed from the beginning that its quantity is negligible small at all times.

[25] J. H. Van Vleck, Paramagnetic relaxation times for titanium and chrome alum, Phys. Rev. **57**, 426 (1940).

[26] R. Orbach, On the theory of spin-lattice relaxation in paramagnetic salts, Proc. Phys. Soc. London **77**, 821 (1961).

[27] A. M. Tyryshkin, S. A. Lyon, A. V. Astashkin, and A. M. Raitsimring, Electron spin relaxation times of phosphorus donors in silicon, Phys. Rev. B **68**, 193207 (2003).

[28] P. Becker, H. J. Pohl, H. Riemann, and N. Abrosimov, Enrichment of silicon for a better kilogram, Phys. Status Solidi A **207**, 49 (2010).

[29] G. C. Bjorklund, Frequency-modulation spectroscopy: A new method for measuring weak absorptions and dispersions, Opt. Lett. **5**, 15 (1980).

[30] A. Yang, M. Steger, T. Sekiguchi, M. L. W. Thewalt, T. D. Ladd, K. M. Itoh, H. Riemann, N. V. Abrosimov, P. Becker, and H.-J. Pohl, Simultaneous Subsecond Hyperpolarization of the Nuclear and Electron Spins of Phosphorus in Silicon by Optical Pumping of Exciton Transitions, Phys. Rev. Lett. **102**, 257401 (2009).

[31] Since the time intervals are long, specially coated 1 and 4 K windows ensure that the infrared 300 K Planck radiation is efficiently blocked. The cryogenic input and output windows are closed completely by aluminum shutters during the pause between each scan to prevent any environmental radiation to cause spin depolarization.

[32] K. Sugihara, Concentration dependent spin-lattice relaxation in n-type silicon, J. Phys. Chem. Solids **29**, 1099 (1968).

[33] Our measured spin relaxation times are much longer compared to Ref. [21]. This difference probably results from the higher sample quality, i.e., a lower concentration of residual, fast spin relaxation centers (see, e.g., Ref. [22] for longer, concentration-dependent spin-lattice relaxation times in natural silicon). The linear temperature dependence of Γ is in this parameter regime consistent with temperature dependent measurements by Honig and Stupp (see Fig. 5 in Ref. [32]). We therefore expect that the maximal $T_1$ becomes even longer at again lower impurity levels.

[34] H. Hasegawa, Spin-lattice relaxation of shallow donor states in Ge and Si through a direct phonon process, Phys. Rev. **118**, 1523 (1960).

[35] A. Morello, J. J. Pla, F. A. Zwanenburg, K. W. Chan, K. Y. Tan, H. Huebl, M. Möttönen, C. D. Nugroho, C. Yang, J. A. Van Donkelaar, A. D. Alves, D. N. Jamieson, C. C. Escott, L. C. Hollenberg, R. G. Clark, and A. S. Dzurak, Single-shot readout of an electron spin in silicon, Nature (London) **467**, 687 (2010).